\documentclass[a4paper,11pt]{article}
\pdfoutput=1
\usepackage{jheppub}
\usepackage[T1]{fontenc}
\usepackage{graphicx}
\usepackage{amsmath}
\usepackage{xspace}
\usepackage{subfig}
\usepackage{xcolor}
\usepackage{mathtools}
\usepackage{braket}
\usepackage{multirow}
\usepackage{colortbl}
\usepackage{setspace}
\usepackage{slashed}
\usepackage{appendix}
\usepackage{booktabs}
\usepackage{cancel}
\usepackage{array}
\usepackage{pst-node} 
\usepackage{tikz}
\usepackage{wasysym}
\usepackage{amssymb}
\usepackage{cleveref}
\usepackage{makecell}
\usetikzlibrary{tikzmark}
\usetikzlibrary{positioning}
\usepackage{hyperref}
\hypersetup{colorlinks=true, linkcolor=[rgb]{0,0.3,0.7}, urlcolor=[rgb]{0,0.3,0.7}, citecolor=[rgb]{0,0.3,0.7}}

\allowdisplaybreaks

\usepackage[detect-all]{siunitx}
\newcommand{\lr}[1]{\left( #1\right)}
\newcommand{\lrs}[1]{\big( #1\big)}
\newcommand{\hc}{\text{h.c.}}
\newcommand{\eq}{\text{eq}}
\newcommand{\im}{\text{Im}}

\newcommand{\tr}{\text{Tr}}
\newcommand{\BR}[1]{\text{BR}\big( #1\big)}

\Crefname{table}{}{}
\Crefname{figure}{}{}
\crefname{equation}{Eq.}{Eqs.}
\crefname{figure}{Fig.}{Figs.}
\crefname{table}{Tab.}{Tabs.}
\definecolor{newgray}{gray}{0.85}
\definecolor{newgray2}{gray}{0.95}
\newcolumntype{P}[1]{>{\centering\arraybackslash}p{#1}} 
\newcolumntype{M}[1]{>{\centering\arraybackslash}m{#1}} 
\newcolumntype{L}[1]{>{\raggedright\arraybackslash}m{#1}} 
\newcolumntype{R}[1]{>{\raggedleft\arraybackslash}m{#1}} 
\newcolumntype{?}{!{\vrule width 0.8pt}}

\title{Leptogenesis and neutrino mass with scalar leptoquarks
}

\author[a,b]{K\r{a}re Fridell}
\emailAdd{karef@post.kek.jp}
\affiliation[a]{\AddrKEK}
\affiliation[b]{\AddrFSU}

\preprint{KEK-TH-2665}

\newcommand{\AddrKEK}{Theory Center, Institute of Particle and Nuclear Studies,\\ High Energy Accelerator Research Organization (KEK), Tsukuba 305-0801, Japan}
\newcommand{\AddrFSU}{Department of Physics, Florida State University, Tallahassee, FL 32306, USA}

\abstract{Leptoquarks are known to generate a wide range of potentially observable phenomena, and have been searched for in different experiments. We show that the observed baryon asymmetry and neutrino mass scale can both be simultaneously produced in a model featuring scalar leptoquarks while avoiding existing experimental constraints and potentially leading to future observable signatures.}

\begin{document}	
\maketitle
\flushbottom

\section{Introduction}
The origin of the observed asymmetry between matter and antimatter in the Universe is a long-standing problem of particle physics. From measurements of the cosmic microwave background (CMB) we know that this asymmetry is relatively small~\cite{Planck:2018vyg},
\begin{equation}
    \eta_B^\text{obs}=(6.20\pm 0.15)\times 10^{-10}
\end{equation}
at $68\%$~C.L., where $\eta_B\equiv n_B/n_\gamma$ is the number density of baryons $n_B$ normalized to that of photons $n_\gamma$, yet how it is generated is unknown. For the asymmetry to be dynamically generated the underlying theory must violate the conservation of baryon number $B$.

Another unsolved problem of particle physics is the origin of neutrino masses $m_\nu$. From observations of neutrino oscillations~\cite{Super-Kamiokande:1998kpq,SNO:2002tuh} we know that at least two neutrino mass eigenstates are non-zero, yet the mechanism behind neutrino mass generation remains unknown. For pure Dirac neutrinos the corresponding Yukawa couplings $y_\nu$ must be very small, $y_\nu\sim 10^{-12}$. Another alternative is that neutrinos are Majorana particles, in which case their corresponding mass term violates the conservation of lepton number $L$. 

Sphaleron interactions violate the conservation of $B+L$, the sum of baryon- and lepton numbers, and are predicted to be active in the early Universe~\cite{Khlebnikov:1988sr}. Any theory that only violates $L$ can therefore indirectly lead to $B$ violation via the sphaleron interactions. This fact, together with the observation that both $\eta_B$ and $m_\nu/\Lambda_\text{EWSB}$ are very small numbers (where $\Lambda_\text{EWSB}$ is the scale of electroweak symmetry breaking), may suggest that the generation of neutrino masses and the baryon asymmetry of the Universe (BAU) could have a common origin in a process known as leptogenesis.

One such scenario is the type-I seesaw model~\cite{Minkowski:1977sc,Yanagida:1979as,Gell-Mann:1979vob,Mohapatra:1980yp}, in which the right-handed neutrinos $N$ have a Majorana mass term $m_{N}\gtrsim 10^9$~GeV~\cite{Davidson:2002qv}. This model can accommodate the observed neutrino mass spectrum without invoking extremely small couplings, and can lead to the generation of a BAU via standard leptogenesis~\cite{Fukugita:1986hr}, i.e.\ the out-of-equilibrium decay of $N$ into a lepton- and Higgs doublet pair. One drawback with this model is that it is difficult to observe experimentally due to the high mass scale of $N$, and the fact that its only interaction with the SM is via the Yukawa coupling to the Higgs. The scale $m_N$ can be lowered e.g.\ in models of resonant leptogenesis~\cite{Pilaftsis:2003gt}, but this relies on a very small mass splitting between the different generations of $N$. 

At the same time, we know that there exists a class of radiative neutrino mass models in which the Majorana masses are generated at e.g.\ 1-loop order~\cite{Zee:1980ai,Tao:1996vb,Ma:1998dn,Ma:2006km} (see Ref.~\cite{Cai:2017jrq} for a review). If the fermion running in the loop is a quark the corresponding bosonic mediator would be a leptoquark~\cite{Chua:1999si,Mahanta:1999xd,AristizabalSierra:2007nf,Dorsner:2017wwn,Cata:2019wbu,Babu:2019mfe,Deppisch:2020oyx,Fajfer:2024uut,Dev:2024tto}. Leptoquark models lead to several distinct observables apart from neutrino mass generation~\cite{Buchmuller:1986zs,Hirsch:1996qy,Hirsch:1996ye}, and can potentially lead to leptogenesis~\cite{Ma:1998hn,Babu:2012vb,Babu:2012iv,Hati:2018cqp,Blazek:2024efd}. However, it has remained unknown whether leptoquarks can simultaneously lead to the generation of neutrino masses and the observed BAU. In Ref.~\cite{Blazek:2024efd} leptoquarks were shown to potentially lead to the leptogenesis via scattering, but the corresponding neutrino mass scale was found to be too large by several orders of magnitude. However, the model in Ref.~\cite{Blazek:2024efd} could still lead to the observed neutrino mass spectrum via cancellations from other neutrino mass contributions.

In this work, we show that neutrino masses and leptogenesis via decay can both be simultaneously generated without special cancellations in a model featuring scalar leptoquarks. This result relies on the consistent treatment of $\Delta L=2$ washout channels, for which we find that the corresponding reaction rate is proportional to the size of the neutrino mass.

In Sec.~\ref{sec:model} we describe the model and neutrino mass generation mechanism, and in Sec.~\ref{sec:lg} we describe the leptogenesis mechanism, while the corresponding details are shown in Appendix~\ref{app:BE}. In Sec.~\ref{sec:pheno} we discuss different phenomenological constraints, and in Sec.~\ref{sec:results} we show our results. We conclude in Sec.~\ref{sec:conclusions}.

\section{The model}\label{sec:model}

The model that we consider consists of an extension of the Standard Model (SM) field content by three scalar leptoquarks: $S_1\in (\bar 3, 1, 1/3)$,  $\tilde R_2\in (3, 2, 1/6)$, and $S_3\in (\bar 3, 3, 1/3)$. The Lagrangian is (c.f.\ Ref.~\cite{Dorsner:2016wpm})
\begin{equation}
	\label{eq:fullLQlagrangian}
	\begin{aligned}
		\mathcal{L} = &\mathcal{L}_{\text{SM}} - \tilde{R}_2^{\dagger}(\Box + m_{\tilde{R}_2}^2)\tilde{R}_{2} - S_1^*(\Box + m_{S_1}^2)S_1 - S_3^a{}^\dagger(\Box + m_{S_3}^2)S_3^a- \lambda_1 v_{B-L}S_1 H^{\dagger}\tilde{R}_{2} \\
       &- \lambda_3 v_{B-L} H^{\dagger}\sigma^a S_3^a\tilde{R}_{2} - \textsl{g}_{2}^{ij}\bar{L}_{i} i\sigma_2\tilde{R}_{2}^{\dagger}d_{j} - \textsl{g}_{1}^{ij}\bar Q^c_i i\sigma_2L_jS_1 -\textsl{g}_{3}^{ij}\bar Q^c_i i\sigma_2\sigma^aS_3^aL_j
  + \text{h.c.}
	\end{aligned}
\end{equation}
Here $\Box=\eta^{\mu\nu}\partial_\mu\partial_\nu$ is the d'Alembert operator. The representations under the SM gauge group $SU(3)_c\times SU(2)_L\times U(1)_Y$ of the different fields, as well as their Lorentz structure, are shown in Tab.~\ref{tab:fields}, where $S$ and $F_{L(R)}$ denote a scalar and left- (right-) handed fermion field, respectively, and the last parenthesis shows $(3B,L)$, where $B$ and $L$ are the baryon- and lepton numbers, respectively. Note that both $S_1$ and $S_3$ can have additional couplings to the SM fermions not included in Eq.~\eqref{eq:fullLQlagrangian}, we have omitted them since they are not directly involved in the leptogenesis mechanism. We have also considered a scenario where the tri-scalar couplings $\lambda_1 v_{B-L}$ and $\lambda_3 v_{B-L}$ are generated by the same $B-L$-breaking vacuum expectation value (VEV) $v_{B-L}$. This was done for simplicity but does not represent the most general scenario. All three leptoquarks are assigned a lepton number $L=-1$, while $S_1$ and $S_3$ have baryon number $B=-1/3$ and $\tilde R_2$ has $B=1/3$. With this assignment the two tri-scalar couplings in Eq.~\eqref{eq:fullLQlagrangian} violate lepton number by two units, $\Delta L=2$, and can therefore potentially lead to Majorana neutrino masses and the generation of a BAU via leptogenesis. 

\begin{table}[]
    \centering
    \begin{tabular}{c|l}
   Field      & Representation\\
   \hline
   $H$ & $S(1,2,1/2)(0,0)$\\
   $L$ & $F_L(1,2,-1/2)(0,1)$\\
   $\bar e^c$ & $F_R(1,1,-1)(0,1)$\\
   $Q$ & $F_L(3,2,1/6)(1,0)$\\
   $\bar u^c$ & $F_R(3,1,2/3)(1,0)$\\
   $\bar d^c$ & $F_R(3,1,-1/3)(1,0)$\\
   \hline
   $S_1$      & $S(\bar 3, 1, 1/3)(-1,-1)$ \\
   $\tilde R_2$      & $S(3, 2, 1/6)(1,-1)$ \\
   $S_3$      & $S(\bar 3, 3, 1/3)(-1,-1)$  \\
    \end{tabular}
    \caption{Field content of the model described by the Lagrangian in Eq.~\eqref{eq:fullLQlagrangian}. The second column shown the representation under the SM gauge group, where $S$ and $F_{L(R)}$ denote a scalar and left- (right-) handed fermion field, respectively. The last parenthesis shows $(3B,L)$ where $B$ is the baryon number and $L$ the lepton number.}
    \label{tab:fields}
\end{table}

We consider the mass hierarchy $m_{S_3}>m_{S_1}\gg m_{\tilde R_2}\gg \Lambda_\text{EWSB}$. At low energies this leads to an effective Lagrangian~\cite{deGouvea:2007qla,Lehman:2014jma,Liao:2016hru,Cirigliano:2017djv,Deppisch:2017ecm,Fridell:2023rtr}
\begin{equation}
    \mathcal{L}_\text{eff}= C_{\bar{d}LQLH1}^{prst}\mathcal{O}_{\bar{d}LQLH1}^{prst}+C_{\bar{d}LQLH2}^{prst}\mathcal{O}_{\bar{d}LQLH2}^{prst}+\hc
\end{equation}
with two dimension-7 $\Delta L=2$ operators
\begin{equation}
\label{eq:ops}
    \mathcal{O}_{\bar{d}LQLH1}^{prst}=\epsilon_{ij}\epsilon_{mn}\lrs{\overline{d_p}L_r^i}\lrs{\overline{Q_s^c}{}^jL_t^m}H^n\,,\quad \mathcal{O}_{\bar{d}LQLH2}^{prst}=\epsilon_{im}\epsilon_{jn}\lrs{\overline{d_p}L_r^i}\lrs{\overline{Q_s^c}{}^jL_t^m}H^n\, .
\end{equation}
These operators will lead to a number of potentially observable low-energy phenomena, as discussed in Sec.~\ref{sec:pheno}. The Wilson coefficients are generated by the model parameters for both $S_1$ and $S_3$. However, since we consider the hierarchy $m_{S_3}>m_{S_1}$, the dominant contribution will come from $S_1$, and we therefore neglect the $S_3$ component of the Wilson coefficients, such that
\begin{equation}
	\label{eq:wilson}
	C^{prst}_{\bar dLQLH1} = -\frac{\lambda_1 v_{B-L} \textsl{g}_{1}^{ps} \textsl{g}_{2}^{rt}}{m_{\tilde{R}_2}^2 m_{S_1}^2}\, , \quad C^{ijkn}_{\bar dLQLH2} = \frac{\lambda_1 v_{B-L} \textsl{g}_{1}^{ps} \textsl{g}_{2}^{rt}}{m_{\tilde{R}_2}^2 m_{S_1}^2}\, .
\end{equation}
\begin{figure}
    \centering
    \includegraphics[width=0.5\linewidth]{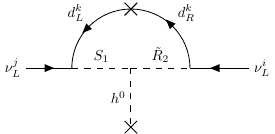}
    \caption{Radiative neutrino mass generated by the model described in the text.}
    \label{fig:mnu}
\end{figure}
Upon electroweak symmetry breaking the component of $\tilde R_2$ with hypercharge $Q=-1/3$ will mix with both $S_1^*$ and the $Q=-1/3$ component of $S_3^\dagger$ via the two tri-scalar couplings to the SM Higgs. However, since we consider the mass hierarchy $m_{S_3}>m_{S_1}$, we expect that the main contribution to the neutrino mass will come from the mixing of $\tilde R_2$ with $S_1$ rather than $S_3$. The presence of $S_3$ is still a crucial component of the model, since without it there would be no $CP$-violation in the decays of $S_1$ during the leptogenesis mechanism (c.f.\ Sec.~\ref{sec:lg}). This role could also be filled by another leptoquark, such as a second copy of $S_1$, and we have here chosen $S_3$ as an example. Note that $S_1$ and $S_3$ can also mix at 1-loop level, but we neglect this effect here due to the smallness of this mixing (c.f.\ Sec.~\ref{sec:lg}). Considering only the $S_1$ contribution we have the neutrino mass matrix~\cite{AristizabalSierra:2007nf,Dorsner:2017wwn,Babu:2019mfe}
\begin{equation}
	\label{eq:LQ1loopmass}
	(m_{\nu})_{ij}=\frac{3\sin\lr{2\theta}v(\textsl{g}_{1}^ T\cdot y^d\cdot \textsl{g}_2+\textsl{g}_2^ T\cdot y^d\cdot \textsl{g}_{1})_{ij}}{32\pi^2}\log\frac{m_{\text{LQ}_1}^2}{m_{\text{LQ}_2}^2}\, ,
\end{equation}
where the mixing angle $\theta$ is given by
\begin{equation}
	\tan(2\theta)=\frac{2\lambda_{1} v_{B-L} v}{m_{\tilde{R}_2}^2-m_{S_1}^2}\, ,
\end{equation}
and the mass eigenstates $m_{\text{LQ}_{1,2}}$ by
\begin{equation}
	m_{\text{LQ}_{1,2}}^2=\frac{1}{2}\lr{m_{\tilde R_2}^2+m_{S_{1}}^2\pm \sqrt{(m_{\tilde R_2}^2-m_{S_{1}}^2)^2+4\lambda_{1}^2v_{B-L}^2v^2}}\, .
\end{equation}
The corresponding radiative neutrino mass diagram is shown in Fig.~\ref{fig:mnu}. Note that a large hierarchy of scales $m_{S_1}\gg m_{\tilde R_2}$ leads to a neutrino mass expression that is mostly independent of the smaller mass $m_{\tilde R_2}$~\cite{Fridell:2024pmw}. Eq.~\eqref{eq:LQ1loopmass} can be diagonalized using the Pontecorvo–Maki–Nakagawa–Sakata (PMNS) matrix $U$ such that $\hat m_\nu = U^\dagger (m_\nu)_{ij} U$, where $\hat m_\nu$ is diagonal. In subsequent sections we neglect $CP$-violation in the lepton mixing by taking $U$ to be real. To find the model parameters that lead to the observed neutrino mass scale $m_\nu\sim \tr (\hat m_\nu)$ we use the central values of the normal ordering mixing angles and mass splittings from \textsc{NuFIT v6.0}~\cite{Esteban:2024eli}. The minimum value for $m_\nu$ is then chosen as the case where the smallest neutrino mass scale $m_1$ vanishes $m_1\to 0$, where $\hat m_\nu = \text{diag}(m_1,m_2,m_3)$. For the largest
allowed neutrino mass scale we use the result $\sqrt{\sum_i |U_{ei}|^2\hat m_{i}^2}<0.45$~eV at 90\% C.L.\ from the KATRIN experiment~\cite{Katrin:2024tvg} and solve for $m_1$. With this method we find the upper limit $m_1< 0.30$~eV. Due to its model-independent nature, we choose to use the ground-based KATRIN constraint rather than the more stringent limit $\tr(\hat m) < 0.12$~eV at 95\% C.L.\ coming from observations of the CMB by Planck~\cite{Planck:2018vyg}.

In addition to the neutrino mass contribution in Eq.~\eqref{eq:LQ1loopmass}, the model could accommodate right-handed neutrinos and subsequently generate a Dirac mass term or an additional Majorana contribution via the type-I seesaw mechanism. In fact, $S_1$ could have tree-level couplings to right-handed neutrinos and down-type quark singlets. We do not include this possibility here.

\section{Leptogenesis}\label{sec:lg}

The model described in Sec.~\ref{sec:model} has the potential to lead to a baryon asymmetry of the Universe if the three Sakharov conditions~\cite{Sakharov:1967dj}
\begin{enumerate}
    \item $B$-violation
    \item $C$- and $CP$-violation\label{it:CCP}
    \item Out-of-equilibrium dynamics\label{it:ooe}
\end{enumerate}
are fulfilled. The SM fermion fields are not chirally symmetric and therefore already violate $C$ symmetry. Furthermore, the SM electroweak sphaleron interactions violate $B+L$ symmetry but rapidly reach equilibrium in the early Universe~\cite{Moore:2000mx,Garbrecht:2014kda} and cannot source a BAU on their own.

\begin{figure}
    \centering
    \includegraphics[height=3cm]{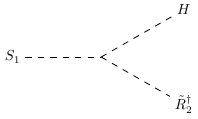}
    \includegraphics[height=3cm]{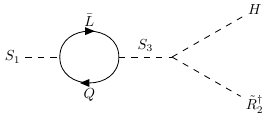}\\
    \caption{Tree-level- and 1-loop decay diagrams leading to $CP$-violation.}
    \label{fig:decay}
\end{figure}

\begin{figure}
    \centering
    \begin{tabular}{cc}
    \includegraphics[height=3cm]{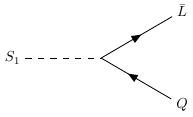} &
    \includegraphics[height=3cm]{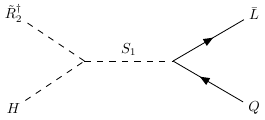}\\
    \includegraphics[height=3cm]{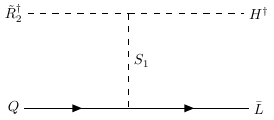} &
    \includegraphics[height=3cm]{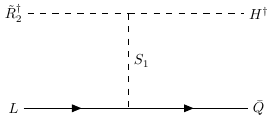}
    \end{tabular}
    \caption{\textbf{Top left:} Decay diagram for $S_1\to\bar L\bar Q$. \textbf{Top right and bottom row:} Scattering diagrams involved in the $\Delta L=2$ washout.}
    \label{fig:diagrams}
\end{figure}

In our scheme the BAU is generated via $\Delta L=2$ out-of-equilibrium $S_1\to H \tilde R_2^\dagger$ decays. We ignore the dynamics of $S_3$ since its mass is greater than that of $S_1$, and its abundance has therefore presumably decayed away before the onset of $S_1$ decays. Note however that $S_3$ is still needed for successful leptogenesis due to it's contribution to loop-level $S_1$ decays, leading to $CP$-violation. We further neglect thermal effects such as thermal masses and thermal corrections to decays, scattering processes, and $CP$-asymmetries (see e.g.\ Ref.~\cite{Giudice:2003jh}). The $CP$-violation in our mechanism comes from the interference between tree- and loop-level decay diagrams of $S_1$ shown in Fig.~\ref{fig:decay}. Note that there is also $CP$-violation in $S_1\to \bar L\bar Q$ decays (see Fig.~\ref{fig:diagrams} top left). Since the total widths of $S_1$ and $S_1^*$ must remain equal for $CPT$ to be conserved, we require that this latter source of $CP$-violation is equal in magnitude but opposite in sign compared to the former. Note that, in this model, there are no 1-loop diagrams with vertex corrections. Note further that 1-loop self-energy diagrams with scalar fields in the loop do not contribute to $CP$-violation, since the corresponding interference term is purely real. The amount of $CP$-violation $\epsilon$ is defined as
\begin{equation}
    \epsilon \equiv \BR{S_1\to H \tilde R_2^\dagger}-\BR{S_1^*\to H^\dagger \tilde R_2}\, ,
\end{equation}
which for the model described in Sec.~\ref{sec:model} and using the diagrams shown in Fig.~\ref{fig:decay} is given by (c.f.\ Refs.~\cite{Babu:2012vc,Chongdar:2021tgm,Fridell:2021gag})
\begin{equation}
\label{eq:eps}
    \epsilon=\frac{2}{\pi}\frac{|v_{B-L}|^2\im\big[\lambda_1\lambda_3^*\tr(\textsl{g}_3^\dagger \textsl{g}_1) \big]}{4\tr(\textsl{g}_1^\dagger \textsl{g}_1)m_{S_1}^2+|\lambda_1v_{B-L}|^2}\frac{x}{1-x}\, ,
\end{equation}
where $x\equiv m_{S_1}^2/m_{S_3}^2$. 

In addition to the decays of $S_1$, the present model will also lead to $\Delta L = 2$ washout processes (see Fig.~\ref{fig:diagrams} top right and bottom row), as well as both $\Delta L = 2$ and $\Delta L = 0$ processes with external $S_1$ legs. Since we neglect thermal effects we also treat all light fields as massless, including $\tilde R_2$ and all SM fields. In this limit the two-to-two scattering processes with external $S_1$ legs are IR divergent, and we therefore neglect them. This is further motivated since we expect that the main source of washout comes from the $S_1$-mediated $\Delta L=2$ processes (without external $S_1$ legs) because they are not suppressed by the small $S_1$ number density during the regime of strong washout. In Sec.~\ref{sec:results} we test the validity of this statement by artificially modifying the corresponding set of equations.

By including the processes shown in Figs.~\ref{fig:decay} and~\ref{fig:diagrams} while taking $\tilde R_2^\dagger \leftrightarrow L+\bar d$ to be in chemical equilibrium we arrive at the following set of Boltzmann equations (c.f.\ Appendix~\ref{app:BE}),
\begin{align}
    zHn_\gamma\frac{d\eta_{S_1^+}}{dz} &=-\gamma_D\left[\frac{\eta_{S_1^+}}{\eta_{S_1}^\eq}-1+\epsilon c_+\frac{\eta_{\Delta (B-L)}}{\eta^\eq_{B-L}}\right]\, \label{eq:be1m},\\
   zHn_\gamma\frac{d\eta_{S_1^-}}{dz} &=-\gamma_D\left[\frac{\eta_{S_1^-}}{\eta_{S_1}^\eq}-c_-\frac{\eta_{\Delta (B-L)}}{\eta^\eq_{B-L}}\right]\, \label{eq:be2m},\\
        \frac{zHn_\gamma}{c_{\Delta(B-L)}}\frac{d\eta_{\Delta (B-L)}}{dz} &= \frac{\gamma_D}{2}\left[r\frac{\eta_{S_1^-}}{\eta_{S_1}^\eq}+\epsilon\left(\frac{\eta_{S_1^+}}{\eta_{S_1}^\eq}-1\right)\right]-\left(c_-r\frac{\gamma_D}{2}+c_+\gamma_W\right)\frac{\eta_{\Delta (B-L)}}{\eta_{B-L}^\eq}\, \label{eq:be3m}.
    \end{align}
Here $z\equiv m_{S_1}/T$ is a time-evolution variable, where $T$ is the temperature, $H$ is the Hubble rate, $n_\gamma$ is the photon number density, and $\eta_X=n_X/n_\gamma$ is the number density of particle $X$ normalized to the photon number density, where a superscript \textit{eq} denotes the equilibrium density and where $\eta_{\Delta(B-L)}$ denotes the net $B-L$ number density. Furthermore, $\gamma_D$ and $\gamma_W$ are the equilibrium reaction rate densities for the decay of $S_1$ and washout, respectively. Furthermore, $r$ is given by
\begin{equation}
    r \equiv \BR{S_1\to H \tilde R_2^\dagger}+\BR{S_1^*\to H^\dagger \tilde R_2}\, ,
\end{equation}
while $c_\pm$ and $c_{\Delta (B-L)}$ are $\mathcal{O}(\text{few})$ numerical coefficients that depend on the number of thermally active fermion species. See Appendix~\ref{app:BE} for a derivation of \Cref{eq:be1m,eq:be2m,eq:be3m} as well as definitions of the different variables.
Lastly, rather than tracking the evolution of $S_1$ and $S_1^*$ we find it more convenient to use $S_1^\pm$ where
\begin{align}
    & \eta_{S_1^+}\equiv \frac{1}{2}(\eta_{S_1}+\eta_{S_1^*})\, ,\\
    & \eta_{S_1^-}\equiv \frac{1}{2}(\eta_{S_1}-\eta_{S_1^*})\, .
\end{align}
Note that in the equation for $\eta_{S_1^-}$ the corresponding equilibrium is reached when its density vanishes, $\eta_{S_1^-}^\eq=0$, while $\eta_{S_1^+}$ reaches equilibrium when the densities for $S_1$ and $S_1^*$ are equal, such that $\eta_{S_1^+}^\eq=\eta_{S_1}^\eq$. The first terms in both Eqs.~\eqref{eq:be1m} and~\eqref{eq:be2m} therefore have the form
\begin{equation}
    zHn_{\gamma}\frac{d\eta_{S_1^\pm}}{dz}\supset -\gamma_D\frac{1}{\eta_{S_1}^\eq}\left(\eta_{S_1^\pm}-\eta_{S_1^\pm}^\eq\right)\, .
\end{equation}
The same form appears in the first two terms of Eq.~\eqref{eq:be3m} although with additional factors $\epsilon$ and $r$. Since $S_1$ is charged under $SU(3)_c$, it can be produced in $gg\to S_1 S_1^*$ where $g$ is a gluon. We expect the rate of this reaction to be rapid in the very early Universe, at $T\gg m_{S_1}$, but quickly fall off around $T\sim m_{S_1}$ due to its dependence on the square of the $S_1$ abundance (in the limit of $\eta_{S_1^-}\to 0$). To account for this interaction we set the initial $S_{1}^+$ and $S_{1}^-$ abundances at $z\ll 1$ to their equilibrium values when we numerically solve the Boltzmann equations in Sec.~\ref{sec:results}.

We note that in \Cref{eq:be1m,eq:be2m,eq:be3m} there is no dependence on the mass of $\tilde R_2$. Since we assume a large hierarchy $m_{S_1}\gg m_{\tilde R_2}$, we treat $\tilde R_2$ as ultra-relativistic during the generation of the asymmetry, i.e.\ at temperatures $0.01\lesssim z\lesssim 100$, and its mass therefore does not enter the equations. Furthermore, in the limit of this hierarchy, the washout diagrams in Fig.~\ref{fig:diagrams} (top right and bottom row) are roughly proportional to the neutrino mass, i.e.\ they have the same dependence on the mass and two couplings of $S_1$ as $m_\nu$ does. What separates them is that $m_\nu$ depends on the Higgs VEV as well as the coupling $\textsl{g}_2$ between $\tilde R_2$ and the SM, whereas the washout diagrams do not. 

In our model the asymmetry is generated by $\Delta L=2$, $\Delta B=0$ interactions, and it would therefore have been enough to track the $L$ asymmetry rather than $B-L$ asymmetry in Eq.~\eqref{eq:be1m}, since all the $B$ violation cancels out. We have chosen to track the $B-L$ asymmetry for generality. However, note that the observed BAU is completely contained in the baryon sector, and to compare our results with observations we use the conversion~\cite{Harvey:1990qw}
\begin{equation}
    Y_{\Delta B}=\frac{79}{51}Y_{\Delta (B-L)}\, .
\end{equation}
An additional factor $d_\text{rec}\approx 1/29$ should also be included to account for the change in entropy density between the early and late Universe. Note that this factor is slightly smaller than in the standard type-I seesaw leptogenesis since we have included the relativistic degrees of freedom of $\tilde R_2$. Furthermore, while a relativistic $\tilde R_2$ can lead to the reaction $\tilde R_2^\dagger \leftrightarrow L+\bar d$ being in chemical equilibrium, this does not modify the sphaleron conversion factor, since the sphaleron only involves the SM fermions. This reaction does however modify the overall chemical equilibrium relations between different particle species, c.f.\ Appendix~\ref{app:BE}.

As mentioned in Sec.~\ref{sec:model}, the leptoquark model considered here could be extended with right-handed neutrinos $N$ that couple to $S_1$ and $\bar d^c$ at tree-level. For Majorana $N$ this could lead to additional channels of $CP$-violation in $S_1$ decays via vertex corrections, and therefore a possible enhancement of the asymmetry~\cite{Babu:2012iv,Babu:2012vb,Hati:2018cqp}. However, if $N$ is lighter than $S_1$ it could potentially lead to additional $N$-mediated $\Delta L=2$ washout channels that erase the asymmetry, depending on its mass and the size of its coupling to the SM. If $N$ does not have a Majorana mass, it could still affect the leptogenesis mechanism as an external state by leading to additional $S_1$-mediated $\Delta L=2$ washout channels. 

\section{Phenomenology}\label{sec:pheno}
The model presented in Sec.~\ref{sec:model} is subject to a number of experimental constraints, as has been discussed in recent works~\cite{Fajfer:2024uut,Dev:2024tto}. Note that each such experimental observable will either depend on the mass of $\tilde R_2$ or rely on couplings that are not present in the leptogenesis mechanism or neutrino mass generation. In the limit of large hierarchy $m_{S_1}\gg m_{\tilde R_2}$, both the neutrino mass and final baryon asymmetry are independent of $m_{\tilde R_2}$, as discussed in Secs.~\ref{sec:model} and~\ref{sec:lg}, respectively. We can therefore always tune $m_{\tilde R_2}$ in such a way that the model is unconstrained in the region of successful leptogenesis and neutrino mass generation, so long as such a region exists while still respecting the hierarchy $m_{S_1}\gg m_{\tilde R_2}$, which will be the case for each observable below. The model is subject to different constraints, depending most significantly on $m_{\tilde R_2}$, and could lead to observable signatures in each phenomenological probe discussed below, but is not required to do so in order for the leptogenesis mechanism to work. For this reason, we comment on various relevant observables in this section but do not attempt to show exclusion lines.

\subsection{Colliders}\label{sec:lhc}
Leptoquarks could be produced in collider experiments such as the LHC~\cite{Blumlein:1996qp,Dorsner:2014axa,Diaz:2017lit,Schmaltz:2018nls,Bandyopadhyay:2018syt,Greljo:2020tgv,Dorsner:2021chv,Crivellin:2021bkd,Bhaskar:2021gsy,Bhaskar:2023ftn,Desai:2023jxh,Varzielas:2023qlb}. For the model we consider here, only $\tilde R_2$ could potentially be within the reach of collider searches, since we take the masses of $S_1$ and $S_3$ to be much greater. The most stringent constraint $m_\text{LQ}>1460$~GeV for $\textsl{g}_2^{33}=3$ at 95\%~C.L.\ comes from a search by ATLAS for pair production of scalar leptoquarks decaying into $b$ and $\tau$~\cite{ATLAS:2023uox}. More stringent constraints could be set in future colliders such as FCC-ee~\cite{FCC:2018evy}.

\subsection{Neutrinoless double beta decay}
Neutrinoless double beta ($0\nu\beta\beta$) decay is he most sensitive low-scale experimental probe of lepton number violation in the first-generation fermions~\cite{Pas:1999fc,Deppisch:2012nb,Rodejohann:2012xd}. The model from Sec.~\ref{sec:model} can lead to $0\nu\beta\beta$ decay via the SMEFT operators from Eq.~\eqref{eq:ops}~\cite{Deppisch:2017ecm,Cirigliano:2017djv,Graf:2018ozy,Cirigliano:2018yza,Deppisch:2020ztt,Fridell:2023rtr,Fajfer:2024uut,Dev:2024tto}. From Ref.~\cite{Fridell:2023rtr} we have the constraints
\begin{equation}
	\label{eq:ovbblimit}
	(C^{1111}_{\bar dLQLH1})^{-1/3} < 2.4\times 10^5~\text{ GeV}\, , \quad (C^{1111}_{\bar dLQLH2})^{-1/3} < 1.4\times 10^5~\text{ GeV}\, ,
\end{equation}
which are based on results from the KamLAND-Zen experiment at 90\% C.L.~\cite{KamLAND-Zen:2022tow}. Using the relations in Eq.~\eqref{eq:wilson} while keeping all couplings equal to one, choosing $v_{B-L}=m_{S_1}$, and keeping $m_{\tilde R_1}= 5$~TeV, we find the constraint $m_{S_1}>5.7\times 10^8$~GeV. Instead choosing ${\textsl{g}}_1^{11}=0.025$, $v_{B-L}=3 m_{S_1}$, and $\lambda_1=10^{-3}$ following benchmark point BM1 in Sec.~\ref{sec:results}, we have $m_{S_1}>4.3\times 10^4$~GeV.

\subsection{Rare kaon decays}\label{sec:kaon}
Rare kaon decays $K\to\pi\nu\nu$ are the second most sensitive probe of lepton number violation induced by dimension-7 SMEFT operators~\cite{Li:2019fhz,Deppisch:2020oyx,Fridell:2023rtr,Gorbahn:2023juq,Buras:2024ewl}, and the most sensitive overall beyond the first generation fermions. The operator $\mathcal{O}_{\bar dLQLH1}^{prst}$ from Eq.~\eqref{eq:ops} can lead to such lepton-number-violating rare kaon decays. Currently the most stringent constraint $(C^{2r1t}_{\bar dLQLH1})^{-1/3} < 2.2\times 10^4~\text{ GeV}$~\cite{Fridell:2023rtr} comes from the charged mode $K^+\to\pi^+\nu\nu$ at the NA62 experiment at 90\% C.L.~\cite{NA62:2020fhy,NA62:2021zjw}. For BM1 with $\textsl{g}_1^{1t}=0.025$ we find $m_{S_1}>32$~GeV. These constraints are based on searches for the SM mode $K^+\to\pi^+\nu\bar\nu$, while a dedicated search for the $\Delta L=2$ mode could be possible in future experiments~\cite{HIKE:2022qra}. The future KOTO-II experiment~\cite{Nanjo:2023xvj} will put even more stringent constraints on the corresponding neutral mode $K_L\to\pi^0\nu\nu$. 

The corresponding rare $B$-meson decay $B\to K\pi\nu\bar\nu$ can also potentially lead to signals of lepton number violation. However, existing constraints on the scale of NP from this decay are less stringent than those coming from kaon decays, in the case where NP couples to all quark flavours equally~\cite{Fridell:2023rtr}. For flavour non-universal NP that dominantly couples to the third generation, $B$ decays could potentially provide the most stringent constraints. A recent measurement at Belle~II~\cite{Belle-II:2023esi} shows an excess of events which could be due to NP~\cite{Fridell:2023ssf}, but would lead to a significant over-production of the neutrino mass in the case of lepton number violation~\cite{Buras:2024ewl}.

\subsection{Baryon number violation}

In Tab.~\ref{tab:fields} the leptoquarks $S_1$, $\tilde R_2$, and $S_3$ are denoted as carrying a non-zero baryon number as well as lepton number. With this assignment, baryon number is conserved in all couplings in the corresponding Lagrangian in Eq.~\eqref{eq:fullLQlagrangian}. However, as mentioned above, $S_1$ and $S_3$ can have diquark couplings as well as leptoquark couplings, i.e.\ they can potentially interact with a pair of quarks at tree-level. This would violate both lepton- and baryon number, $\Delta B = -\Delta L =1$. If present, such couplings could potentially lead to rapid proton decay at tree-level~\cite{Gu:2011pf,Dorsner:2012nq,Babu:2012vb,Babu:2012iv,Hati:2018cqp}. We can predict the rate of proton decay in our model, e.g.\ via the mode $p\to\pi^0 e^+$, using the nucleon form factor~\cite{Aoki:2017puj}
\begin{equation}
\label{eq:formfactorPion}
    \bra{\pi}\epsilon_{abc}\left(q^aCP_\Gamma q^b\right)P_{\Gamma'}q^c\ket{\psi} = P_{\Gamma'}\Big[W_0^{\Gamma\Gamma'}(\mu,p^2) -\frac{i\slashed p}{m_\psi}W_1^{\Gamma\Gamma'}(\mu,p^2)\Big]u_\psi\, ,
\end{equation}
where $W_0^{\Gamma\Gamma'}$ and $W_1^{\Gamma\Gamma'}$ are the form factors, $\mu$ is the characteristic energy scale, $p$ is the momentum exchange, $u_\psi\in \{n,p\}$ and $q\in \{u,d\}$ are nucleon- and quark spinors, respectively, $C$ is the charge conjugation operator, $P_{\Gamma}$ is a projection operator for chiral indices $\Gamma,\Gamma'\in\{L,R\}$, and $a,b,c$ are color indices. Considering the coupling of $S_1$ to doublet quarks such that $\mathcal{L}\supset -\textsl{g}_1'^{ij}\bar Q_i i\sigma_2 Q^c_j S_1+\hc$ we have the corresponding matrix element
\begin{equation}
    \mathcal{M}_{p\to\pi^0 e^+}=\bra{\pi^0e^+}C_{Q^3L}^{1111}(\mu_\text{NP})\mathcal{O}_{Q^3L}^{1111}\ket{p} \, ,
\end{equation}
for the operator
\begin{equation}
    \mathcal{O}_{Q^3L}^{prst}=\epsilon_{abc}\epsilon_{ij}\epsilon_{mn}(\overline{Q^c_p}{}^{ai}Q^{bj}_r)(\overline{Q^c_s}{}^{cm}L^n_t)
\end{equation}
and Wilson coefficient
\begin{equation}
    C_{Q^3L}^{prst}=\frac{\textsl{g}_1'^{pr}\textsl{g}_1^{st}}{m_{S_1}^2}\, .
\end{equation}
Here we have neglected the contribution from $S_3$ since we expect $S_1$ to dominate. We then have
\begin{equation}
\mathcal{M}_{p\to\pi^0e^+}= U'(\mu_\text{NP},\mu_0)C_{Q^3L}(\mu_\text{NP})\Big[W_0^{LL}(\mu_0,m_e^2) -\frac{i\slashed p}{m_\psi}W_1^{LL}(\mu_0,m_e^2)\Big]u_pP_L\bar u_{e} \, ,
\end{equation}
where $\mu_0$ is the energy scale of the decay, $\mu_\text{NP}\sim m_{S_1}$ is the scale of New Physics, and
\begin{equation}
    U'(\mu_\text{NP},\mu_0)= U_i^{N_f=6}(\mu_\text{NP},m_t) U_i^{N_f=5}(m_t,m_b) U_i^{N_f=4}(m_b,\mu_0)
\end{equation}
accounts for the running of the strong coupling constant $\alpha_S$, where~\cite{Aoki:2008ku}
\begin{equation}
	\label{eq:runningfactor}
	U_i^{N_f}(\mu_1,\mu_2) = \lr{\frac{\alpha_S(\mu_2)}{\alpha_S(\mu_1)}}^{\gamma_i^{0}/2\beta_0}\Bigg[1+\lr{\frac{\gamma_1}{2\beta_0}-\frac{\beta_1\gamma_0}{2\beta_0^2}}\frac{\alpha_S(\mu_2)-\alpha_S(\mu_1)}{4\pi}\Bigg]\, ,
\end{equation}
for the coefficients
\begin{align}
    &\beta_0=11-\frac{2}{3}N_f,\; \beta_1=102-\frac{38}{3}N_f,\\
    &\gamma_0=-4,\; \gamma_1=-\left(\frac{14}{3}+\frac{4}{9}N_f\right)\, ,
\end{align}
where $N_f$ is the number of fermions that carry colour charge. 

Currently the most stringent constraints on the $p\to\pi^0e^+$ lifetime is $\tau_{p\to\pi^0e^+}>1.6\times 10^{34}$ years at 90\% C.L.\ coming from the Super-Kamiokande experiment~\cite{Super-Kamiokande:2016exg}. Using the form factors given in Ref.~\cite{Aoki:2017puj} we find $C_{Q^3L}^{1111}\lesssim(2.1\times 10^{13}\text{ GeV})^{-2}$. Choosing the parameters $m_{S_1}=10^8$~GeV and $\textsl{g}_{1}^{11}=0.025$ following benchmark point BM1 in Sec.~\ref{sec:results} we then find that we require $\textsl{g}_1'^{11}\lesssim 0.9\times 10^{-9}$ in order to satisfy the experimental bound on proton decay. This is a very small coupling, however there are several caveats which should be considered at this point. Firstly, the neutrino mass generation mechanism depends chiefly on the couplings to the third generation down-type quark, while for this proton decay mode only the first generation couplings are relevant at tree-level. Due to loop suppression and the smallness of the $V_{td}$ and $V_{ub}$ CKM matrix elements, loop-level proton decays that utilize the quark couplings to third generation quarks have a negligible decay width. In a scenario where leptoquarks couple mainly to the third generation, the proton decay constraint could therefore more easily be satisfied. Secondly, if one only considers leptogenesis without requiring that the model also leads to the observed neutrino mass, it is possible to have much higher masses for $S_1$ and $S_3$ (c.f.\ Sec.~\ref{sec:results}), which would significantly increase the proton lifetime. This would underproduce the neutrino mass, and a different mechanism would then be required for its successful generation (e.g.\ type-I seesaw). Lastly there could be cancellations between different contributions to the decay width of the proton coming from e.g.\ $S_3$ or other components of a more complete model. 

Note that with $\textsl{g}_1'$ we could also have the proton decay modes $p\to \pi^+ \nu$ mediated by a dimension-7 operator involving $\tilde R_2$-$S_1$ mixing. This mode has a less stringent bound compared to that of $p\to\pi^0e^+$, namely $\tau_{p\to\pi^+\nu}>3.9\times 10^{32}$ years at 90\% C.L.\ again from the Super-Kamiokande experiment~\cite{Super-Kamiokande:2013rwg}. The corresponding decay rate will depend on $m_{\tilde R_2}$ in addition to $\textsl{g}_1'$, making it two parameters that are not involved in the leptogenesis or neutrino mass generation mechnaisms in the $m_{S_1}\gg m_{\tilde R_2}$ limit.

As mentioned in Secs.~\ref{sec:model}~and~\ref{sec:lg}, the leptoquark model could be extended with right-handed neutrinos $N$. If $N$ has a sizable Majorana mass, its coupling to $S_1$ and $\bar d^c$ could lead to potentially observable neutron-antineutron ($n\text{\--}\bar n$) oscillation~\cite{Kuo:1980ew,Mohapatra:1980qe,Babu:2008rq}. For this mechanism to work, $S_1$ additionally needs diquark couplings to a pair of quark doublets or up- and down-type singlets, same as for proton decay. The corresponding $n\text{\--}\bar n$ oscillation diagram then consists of two $S_1$ mediators and one $N$ mediator with six external quark legs.

\subsection{Flavour violation}\label{sec:flavour}
For flavour off-diagonal couplings there are several observables of flavour physics that could constrain the leptoquark model~\cite{Fajfer:2024uut,Dev:2024tto}, e.g.\ charged lepton-flavour-violating processes such as $\mu\to e\gamma$, $\mu N\to e N$ (where $N$ is a nucleus), $K_L\to \mu^\pm e^\mp$, and $K^+\to \pi^+\mu^+ e^-$, or deviations from the SM in quark-flavour observables such as $K\to\pi\nu\bar\nu$. Note that this latter observable is distinct from the lepton-number-violating process $K\to\pi\nu\nu$ discussed in Sec.~\ref{sec:kaon}. In our present scenario the leptoquark $\tilde R_2$ is the field that is potentially susceptible to such constraints since it is the lightest.

From Refs.~\cite{Fajfer:2024uut} and~\cite{Dev:2024tto} we find that the most stringent such constraint comes from $\mu N\to eN$ conversion using gold at the SINDRUM~II experiment, namely $\text{BR}(\mu N \to e N)_\text{Au}<7\times 10^{-13}$ at 90\% C.\ L.~\cite{SINDRUMII:2006dvw}, where BR here denotes the rate of conversion normalized to the muon capture rate. We then have the constraint $\textsl{g}_{2}^{11}\textsl{g}_{2}^{21}/m_{\tilde R_2}^2<(500\text{ TeV})^2$~\cite{Fajfer:2024uut}.

For a completely general flavour structure we see that constraints coming from flavour violation are more stringent than those coming from collider searches (c.f.\ Sec.~\ref{sec:lhc}). However, note that neutrino mass generation most dominantly depends on the third generation couplings, for which the flavour constraints are not applicable. In Sec.~\ref{sec:results} we therefore choose to evaluate benchmark points with $\tilde R_2$ close to collider constraints, and neglect the constraints coming from flavour physics.

\section{Results}\label{sec:results}

To find the model parameters that lead to the successful generation of a BAU and neutrino masses that both match observations, we numerically solve the Boltzmann equations from \Cref{eq:be1m,eq:be2m,eq:be3m} and overlap these results with the neutrino mass from Eq.~\eqref{eq:LQ1loopmass}. We define a benchmark point BM1 as
\begin{center}
\begin{tabular}{|cl|}
     \hline 
     \multirow{2}*{BM1:} & $m_{S_1}=10^8$~GeV, $m_{\tilde R_2}=5$~TeV, $m_{S_3}=3m_{S_1}$, $v_{B-L}=3m_{S_1}$,\\
     & ${\textsl{g}}_1=0.025$, ${\textsl{g}}_2 = 1$, ${{\textsl{g}}}_3={\textsl{g}}_1$, $\lambda_1=10^{-3}$, $\lambda_3=i\lambda_1$\\
     \hline
\end{tabular}
\end{center}
leading roughly to the observed neutrino mass. Note that we take $\lambda_3$ to be purely imaginary. We keep these relations between the parameters related to $S_1$ and $S_3$ in order for our approximations to always be valid, i.e.\ that we can neglect the dynamics of $S_3$ etc., such that below when we vary e.g.\ $\textsl{g}_1$ we also vary $\textsl{g}_3$. We consider only third generation couplings $\textsl{g}_i^{mn}\propto \delta_{m3}\delta_{n3}$ for $i\in\{1,2,3\}$ since these are the couplings most relevant for neutrino mass generation. We do not attempt to exactly reproduce the observed neutrino mass spectrum, only its characteristic scale. We furthermore assume that there are three thermally active fermion generations such that $N=3$ (c.f.\ Appendix~\ref{app:BE}), in order to be consistent in the low-$m_{S_1}$ parameter space regions. We find that varying $N$ generally has an $\mathcal{O}(1)$ effect on the results.

\begin{figure}
    \centering
    \includegraphics[scale=1.8]{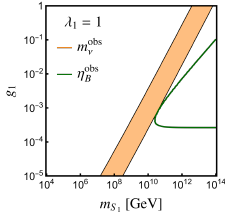}
    \includegraphics[scale=1.8]{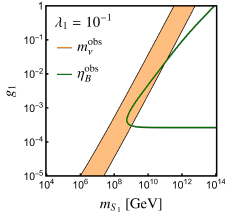}\\[4mm]
    \includegraphics[scale=1.8]{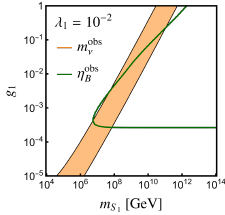}
    \includegraphics[scale=1.8]{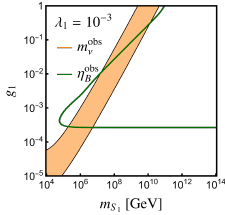}
    \caption{Contour plots in the ${\textsl{g}}_1$-$m_{S_1}$ plane showing the regions in which the observed value for $\eta_B$ (green) and $m_\nu$ (orange) is produced, for different values of the coupling $\lambda_1$ and indicated in the top-left corner of each plot. Other parameters are chosen according to benchmark point BM1 (see text).}
    \label{fig:contour-lam}
\end{figure}

In Fig.~\ref{fig:contour-lam} we show regions in the $\textsl{g}_1$-$m_{S_1}$ plane that lead to the observed value for $\eta_B$ (green) and $m_\nu$ (orange), for the parameters defined as BM1 but varying $\textsl{g}_1$, $m_{S_1}$, and $\lambda_1$. The lower edge of the $m_\nu$ band corresponds to $m_1=0$, while the upper edge corresponds to the maximum value of $m_1<0.30$~eV (c.f.\ Sec.~\ref{sec:model}). We see that there are regions of parameter space in which these regions overlap, such that the model simultaneously leads to both the observed neutrino mass scale as well as the observed BAU. Note that the green region appears as a line that carves out an area towards the right. Within this region (to the right) the generated asymmetry is larger than that which is observed, and outside it it is smaller. For smaller couplings $\lambda_1$ this region grows, since the washout coming from $\Delta L=2$ scatterings is decreased. However, if $\lambda_1$ is decreased to smaller values than the ones shown in Fig.~\ref{fig:contour-lam}, e.g.\ for $\lambda_1\lesssim 10^{-4}$, the green line disappears completely. The reason for this is that for small $\lambda_1$ the $CP$-violation also decreases with $\lambda_1$, such that for small enough couplings the observed BAU can no longer be generated. Similar effects can be seen in the variation of $\textsl{g}_1$, for small enough values $\eta_B$ becomes smaller than $\eta_B^\text{obs}$, independently of the value of $\lambda_1$. Whichever of $\textsl{g}_1$ and $\lambda_1$ is smaller is the one that governs the size of the $CP$-violating parameter $\epsilon$, as can be seen from the expression in Eq.~\eqref{eq:eps}. For large enough couplings $\textsl{g}_1$ or $\lambda_1$ the value of $\eta_B$ is again too small, due to increased $\Delta L=2$ washout, as can be seen in the top-left part of each plot. Note that the $\eta_B^\text{obs}$-line in this region is almost parallel to that of $m_\nu^\text{obs}$. The reason for this is that the $\Delta L=2$ washout processes and $m_\nu$ both have the same proportionality to $m_{S_1}$, $\textsl{g}_1$, and $\lambda_1$ in the limit of large hierarchy, $m_{S_1}\gg m_{\tilde R_2}$. Having a large neutrino mass will therefore also lead to having a large washout.

\begin{figure}
    \centering
    \includegraphics[scale=1.8]{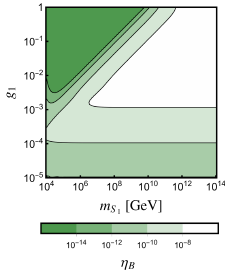}
    \includegraphics[scale=1.8]{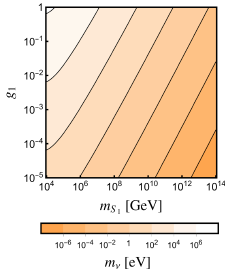}
    \caption{Contour plots in the ${\textsl{g}}_1$-$m_{S_1}$ plane showing the value of $\eta_B$ (left) and $m_\nu$ (right). Other parameters are chosen according to benchmark point BM1 (see text).}
    \label{fig:contour-bench}
\end{figure}

In Fig.~\ref{fig:contour-bench} we show the values of $\eta_B$ (left) and $m_\nu$ (right) for the parameters defined as BM1 but while varying $\textsl{g}_1$ and $m_{S_1}$. We see that $\eta_B$ decreases proportionally to the square of $\textsl{g}_1$ for small values of $\textsl{g}_1$, as expected from Eq.~\eqref{eq:eps}. In contrast, the decrease in $\eta_B$ with increasing $\textsl{g}_1$ in the large-washout regime (top left) is comparatively rapid. In Fig.~\ref{fig:contour-bench} (left) we do not resolve values of $\eta_B$ smaller than $10^{-15}$. 

\begin{figure}
    \centering
    \includegraphics[width=0.49\linewidth]{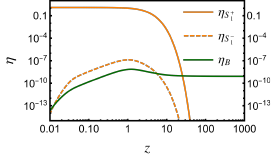}
    \includegraphics[width=0.49\linewidth]{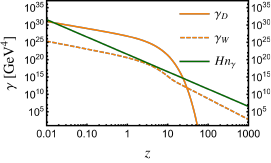}
    \caption{Evolution of $\eta_{S_1^+}$, $\eta_{S_1^-}$, and $\eta_{B}$ with respect to $z\equiv m_{S_1}/T$ (left), as well as the Hubble rate $H$ multiplied by the photon number density $n_\gamma$ and corresponding reaction rate densities $\gamma_D$ and $\gamma_W$ (right), for the parameters described as benchmark point BM1 in the text.}
    \label{fig:line}
\end{figure}

In Fig.~\ref{fig:line} (left) we show the time-evolution of $\eta_{S_1^+}$, $\eta_{S_1^-}$, and $\eta_{B}$ (c.f.\ Sec.~\ref{sec:lg}) with respect to $z\equiv m_{S_1}/T$ for the set of parameters defined as BM1. We see that the $S_1^+$ abundance starts falling off around $z\sim 1$ and vanishes well before $z=100$. The $S_1^-$ abundance increases as $\eta_B$ grows but similarly falls off for $z\gtrsim 1$. The baryon asymmetry $\eta_B$ initially grows as $S_1$ decays, then gets slightly reduced around $1\lesssim z\lesssim 10$ due to the $\Delta L=2$ washouts, after which it freezes out. Note that we here show the value of $\eta_B$ after having applied the sphaleron conversion factor and dilution factor $d_\text{rec}$ (c.f.\ Sec.~\ref{sec:lg}), rather than showing the $B-L$ asymmetry $\eta_{\Delta B-L}$, in order for easier comparison with the observed value $\eta_{B}^\text{obs}\approx 6.20\times 10^{-10}$. In Fig.~\ref{fig:line} (right) we show the corresponding reaction rate densities $\gamma_D$ and $\gamma_W$, as well as the Hubble rate $H$ multiplied by the photon number density $n_\gamma$. When the rates are greater than $Hn_\gamma$ they significantly affect the relevant particle number densities, and while they are smaller they do not have a large effect. Comparing with Fig.~\ref{fig:line} (left) we see that $S_1^+$ decays are significant while $\gamma_D$ is large, which is the era when $\eta_B$ is generated. Around $1\lesssim z\lesssim 10$ the washout rate approaches $Hn_\gamma$, which leads to a small dip in $\eta_B$. If the washout rate were to overtake the expansion, the asymmetry would be significantly reduced, if not removed completely.  

In Sec.~\ref{sec:lg} we neglected the effects of $\Delta L=0$ and $\Delta L = 2$ two-to-two processes with external $S_1$ legs. The $\Delta L=0$ processes can have the effect of reducing the $S_1$ abundance during the temperature regime in which the out-of-equilibrium decays occur, and the $\Delta L=2$ processes lead to additional washout channels. We can estimate the effects of such processes by including artificial terms proportional to $\gamma_D(\gamma_W/\gamma_D)|_{z=1}(\eta_{S_1^+}/\eta_{S_1}^\eq)$ in the first Boltzmann equation, Eq.~\eqref{eq:be1m}, to account for the $S_1$ depletion, and the same but with an additional factor $\eta_{\Delta(B-L)}/\eta_{B-L}^\eq$ in the third Boltzmann equation, Eq.~\eqref{eq:be3m}, to account for the additional washout channels. Here the factor $\gamma_D(\gamma_W/\gamma_D)|_{z=1}$ is included since we expect the processes to have a similar magnitude to that of $\gamma_W$ but a thermal profile similar to that of $\gamma_D$. Doing this we find at most an $\sim\mathcal{O}(1)$ reduction of the final asymmetry for the relevant parameter space.

In Figs.~\ref{fig:contour-lam} and~\ref{fig:contour-bench} the $\tilde R_2$ mass was chosen as $m_{\tilde R_2}=5$~TeV, while the $m_{S_1}$-axis goes down to $10^4$~GeV. Towards this edge the validity of the approximation $m_{S_1}\gg m_{\tilde R_2}$ starts becoming unfounded. The neutrino mass calculation takes $m_{\tilde R_2}$ into account, but for $\eta_B$ we assumed it to vanish, $m_{\tilde R_2}\to 0$. We suspect that having non-zero $m_{\tilde R_2}\lesssim m_{S_1}$ would increase the value of $\eta_B$ in the limit $m_{\tilde R_2}\to m_{S_1}$ since the $\Delta L=2$ washouts become thermally suppressed, so long as the phase space suppression in $S_1$ decays is not too large. 

Furthermore, while $m_\nu$ is independent of $m_{\tilde R_2}$ in the limit $m_{S_1}\gg m_{\tilde R_2}$, we note that part of the successful parameter space in Fig.~\ref{fig:contour-lam} lies in regions where $m_{S_1}$ is only one or a few orders of magnitude greater than $m_{\tilde R_2}$, for which $m_\nu$ is somewhat dependent on $m_{\tilde R_2}$, such that increasing $m_{\tilde R_2}$ would move the orange band towards the top-left corner in each plot. Taking $m_{\tilde R_2}=500$~TeV to agree with the first- and second generation coupling constraints from flavour physics (c.f.\ Sec.~\ref{sec:flavour}) we find that we have an overlap of $m_\nu^\text{obs}$ and $\eta_B^\text{obs}$ only for smaller values of $\lambda_1$.

In this analysis we have not reproduced the neutrino mass spectrum, only its characteristic scale. The full spectrum can be accommodated by using experimental values for the mass splittings $\Delta m_{12}^2$ and $\Delta m_{13}^2$ (in normal hierarchy) and solving Eq.~\eqref{eq:LQ1loopmass} for $\textsl{g}_1$ and $\textsl{g}_2$ for an unknown lightest neutrino mass $m_1$. Between the minimum value $m_1=0$ and highest allowed value $m_1\approx 0.30$~eV (c.f.\ Sec.~\ref{sec:model}) the shape of the diagonalized neutrino mass matrix $\hat m_\nu$ varies greatly. For $m_1=0$ the neutrino mass matrix $\hat m_\nu$ only has non-zero entries in the $(m_1)_{22}$ and $(m_1)_{33}$ components, where the latter is almost and order of magnitude greater than the former. For the maximum value of $m_1$ however, $\hat m$ is close to being proportional to the unit matrix. In this case the hierarchy of leptoquark couplings to the SM fermions will need to be inverted compared to that of the SM Yukawa couplings, on order for one to compensate for the other (c.f.\ Eq.~\eqref{eq:LQ1loopmass}). This is in conflict with the benchmark point used for the leptogenesis mechanism in this section, however note that, neglecting spectator effects, the leptogenesis mechanism we consider here is essentially flavour-blind, such that the same results would apply for $\textsl{g}_2^{(11)}$ as $\textsl{g}_2^{(33)}$.

\section{Conclusion}\label{sec:conclusions}

We have shown that the leptoquark model described in Sec.~\ref{sec:model} can lead to both neutrino mass generation and leptogenesis while avoiding existing experimental constraints. For a small enough mass of $\tilde R_2$ and large enough couplings to the SM, the model can potentially be within the reach of future experiments such as e.g.\ 0$\nu\nu\beta$ decay searches or at colliders. Note however that the leptogenesis mechanism, and to some extent also neutrino mass generation, are decoupled from the mass of $\tilde R_2$.

The leptoquark model was essentially chosen as an example, following these results we may expect a similar situation to appear in other radiative neutrino mass models, or in the inverted hierarchy $m_{\tilde R_2}\gg m_{S_1}$, i.e.\ it is probably possible to have both neutrino masses and leptogenesis in a wide range of scenarios, such as the different UV-completions of 4-fermion $\Delta L=2$ dimension-7 SMEFT operators~\cite{Angel:2012ug,Cai:2014kra,Cepedello:2017eqf,deBlas:2017xtg,Herrero-Garcia:2019czj,Banerjee:2020jun,DasBakshi:2021xbl,Chala:2021juk,Fridell:2022wbz}. The reason we may expect this is that the detailed properties of the new particles, such as Lorentz structure, representation under the SM gauge group, and number of possible tree-level interactions with the SM, generally enter as $\mathcal{O}(\text{few})$ factors in the reaction rates involved in the Boltzmann equations, as well as in the neutrino mass generation mechanism, while the topologies remain the same or very similar. For models where an up-type quark participates in the radiative neutrino mass diagram there may be an even greater successful parameter space available due to the large mass of the top quark. However, these are speculations, and a full study would be warranted to confirm whether such statements are true.

We found that the $S_1$-mediated $\Delta L=2$ washout processes are proportional to the neutrino mass, such that a large neutrino mass would also produce a large washout, leading to the erasure of any generated baryon asymmetry. The smallness of the neutrino mass could then potentially be explained from anthropic arguments (see e.g.\ Ref.~\cite{Adams:2019kby}).

We have neglected thermal effects in the leptogenesis mechanism. This led us to drop some sub-leading washout diagrams which a quick estimate showed could have small but non-vanishing effect, c.f.\ Sec.~\ref{sec:results}. We have furthermore neglected corrections from bound state formation~\cite{Becker:2024vyd}, which can be important for particles charged under a strongly coupled gauge group such as $SU(3)_c$ of the SM. However, bound state formation is unlikely to be significant in the low-$z$ regions during which the BAU is generated in most of our successful parameter space.

\section*{Acknowledgements}
I would like to thank Motoi Endo, Chandan Hati, and Peter Maták for helpful discussions. I acknowledge support from the Japan Society for the Promotion of Science (JSPS)
Grant-in-Aid for Scientific Research B (No. 21H01086 and 23K20847).

\appendix
\section{Boltzmann equations}\label{app:BE}
In this Appendix we give a detailed derivation of the Boltzmann equations from Sec.~\Ref{sec:lg} following Refs.~\cite{Giudice:2003jh,Fridell:2021gag}. The time-evolution of a particle's number density can be written as
\begin{equation}
	zHn_{\gamma}\frac{d\eta_X}{dz}=-\sum_{i,j,\dots}[X\cdots\leftrightarrow ij\cdots]\, .
	\label{eq:boltzmann}
\end{equation}
Here $X$ denotes the particle species, $z\equiv m_X/T$ is the time variable, where $m_X$ is the mass of $X$ and $T$ is the temperature, and $\eta_X=n_X/n_\gamma$ is the number density of $X$ normalized to the photon number density $n_\gamma$ given by
\begin{equation}
    \label{eq:ngamma}
    n_\gamma = \frac{2\zeta(3)}{\pi^2}T^3\, ,
\end{equation}
where $\zeta(3)\approx 1.20$ is the Riemann zeta function. Furthermore, $H$ is the Hubble rate given by
\begin{equation}
	H =\frac{ 1.66\sqrt{g_*}}{m_{\text{Pl}}}T^2\, 
\end{equation}
where $m_\text{Pl}\approx 1.2\times 10^{19}\text{ GeV}$ is the Planck mass and $g^*$ is the number of relativistic degrees of freedom, where in the early Universe our model leads to $g^*= 112.75$, which is greater than in the SM due to the presence of a relativistic $\tilde R_2$. Note that we have here ignored the effects of a relativistic $\tilde R_2$ which could add six degrees of freedom. The square brackets on the RHS of Eq.~\eqref{eq:boltzmann} are given by 
\begin{equation}
	\label{eq:squarebracket}
	[X \cdots \leftrightarrow i\, j \cdots ]=\frac{\eta_X \cdots}{\eta_X^{\text{eq}} \cdots} \gamma^\eq (X \cdots \to i\, j \cdots)-\frac{\eta_i \eta_j \cdots}{\eta_i^{\text{eq}} \eta_j^{\text{eq}} \cdots} \gamma^\eq (i\, j \cdots \to X \cdots)\, .
\end{equation}
Here $i$ and $j$ are additional particles that $X$ interact with, and the dots denote that more particle species can be included in both the initial and final states. The summation in Eq.~\ref{eq:boltzmann} goes over all such interactions that involve $X$. Furthermore, $\eta_X^\eq$ is the normalized equilibrium number density of $X$ given by
\begin{equation}
	\eta_X^\eq=\frac{g_X}{4\zeta(3)}z^2K_2\left(z\right)\, ,
\end{equation}
where $g_X$ is the number of degrees of freedom of $X$ and $K_\nu(z)$ is the modified Bessel function of the second kind. Note that the equilibrium number densities of a given particle and its corresponding antiparticle are equal. The equilibrium reaction rate density $\gamma^\eq$ for a two body decay $X\to ij$ is given by
\begin{equation}
    \gamma^\eq\lr{X \to i\, j}= \eta_{X}^{\text{eq}}n_\gamma\frac{K_1\lrs{z}}{K_2\lrs{z}}\Gamma\lr{X \to i\, j} \, ,
\end{equation}
where
\begin{equation}
	\label{eq:decaywidth}
	\Gamma\lr{X \to i\, j} = \frac{1}{1+\delta_{ij}}\frac{m_{X}^2-m_{i}^2-m_{j}^2}{16\pi m_{X}^3}|\mathcal{M}\lr{X \to i\, j}|^2\, 
\end{equation}
is the decay width for a corresponding matrix element $\mathcal{M}\lr{X \to i\, j}$. For a two-to-two scattering $X a \to i j$ the equilibrium reaction rate density is given by
\begin{equation}
	\gamma^\eq (X\, a \to i\, j) = \frac{T}{64\pi^4}\int_{s_{\min}}^{\infty}ds \sqrt{s}\hat{\sigma}(X\, a \to i\, j)K_1\lr{\frac{\sqrt{s}}{T}}\, ,
\end{equation}
where $s_{\min}=\max\big[{(m_X+m_a)^2,(m_i+m_j)^2}\big]$. Here $\hat{\sigma}$ is the reduced cross section given by
\begin{equation}
\label{eq:sigmahat}
\hat{\sigma}(X\, a \to i\, j) = \frac{1}{8\pi s}\int_{t^-}^{t^{+}}dt|\mathcal{M}(X\, a \to i\, j)|^2\, ,
\end{equation}
with the integration limits
\begin{equation}
	\begin{aligned}
		t^\pm = &\frac{1}{4s}(m_X^2-m_a^2-m_i^2+m_j^2)^2-\frac{1}{4s}\lr{\lambda^{1/2}(s,m_X^2,m_a^2)\mp\lambda^{1/2}(s,m_i^2,m_j^2)}^2\, \,
	\end{aligned}
\end{equation}
where $\lambda(x,y,z)=x^2+y^2+z^2-2xy-2xz-2yz$ is the Källén function. 

To apply this formalism to the leptoquark model given in Sec.~\ref{sec:model} we make the replacement $X\to S_1, S_1^*$, and identify the relevant decay- and scattering processes as those shown in Figs~\ref{fig:decay} and~\ref{fig:diagrams}. Using Eq.~\ref{eq:boltzmann} we then have
\begin{align}
    zHn_\gamma\frac{d\eta_{S_1}}{dz}=-D-D_0\, ,\\
    zHn_\gamma\frac{d\eta_{S_1^*}}{dz}=-\overline{D}-\overline{D}_0\, ,
\end{align}
where
\begin{align}
   & D\equiv \gamma_D\left[(r+\epsilon)\frac{\eta_{S_1}}{\eta_{S_1}^\eq}-(r-\epsilon)\frac{\eta_H\eta_{\tilde R_2^\dagger}}{\eta^\eq_H\eta^\eq_{\tilde R_2^\dagger}}\right]\, ,\\
  &  D_0\equiv \gamma_D\left[(1-r-\epsilon)\frac{\eta_{S_1}}{\eta_{S_1}^\eq}-(1-r+\epsilon)\frac{\eta_{\bar L}\eta_{\bar Q}}{\eta^\eq_L\eta^\eq_Q}\right]\, ,\\
  &  \overline{D}\equiv \gamma_D\left[(r-\epsilon)\frac{\eta_{S_1^*}}{\eta_{S_1}^\eq}-(r+\epsilon)\frac{\eta_{H^\dagger}\eta_{\tilde R_2}}{\eta^\eq_H\eta^\eq_{\tilde R_2^\dagger}}\right]\, ,\\
  &  \overline{D}_0\equiv \gamma_D\left[(1-r+\epsilon)\frac{\eta_{S_1^*}}{\eta_{S_1}^\eq}-(1-r-\epsilon)\frac{\eta_{L}\eta_{Q}}{\eta^\eq_L\eta^\eq_Q}\right]\, .
\end{align}
Here $r$ and $\epsilon$ are given in terms of the branching ratios of $S_1$ and $S_1^*$ as
\begin{align}
    r \equiv \BR{S_1\to H \tilde R_2^\dagger}+\BR{S_1^*\to H^\dagger \tilde R_2}\\
    \epsilon \equiv \BR{S_1\to H \tilde R_2^\dagger}-\BR{S_1^*\to H^\dagger \tilde R_2} \, 
\end{align}
and the total decay rate $\gamma_D$ is given by
\begin{equation}
    \gamma_D \equiv \gamma^\eq_\text{tree}(S_1\to H\tilde R_2^\dagger)+\gamma^\eq_\text{tree}(S_1\to \bar L\bar Q)\, ,
\end{equation}
where $\gamma^\eq_\text{tree}$ is the tree-level decay rate. 
For convenience we define 
\begin{align}
    & \eta_{S_1^+}\equiv \frac{1}{2}(\eta_{S_1}+\eta_{S_1^*})\, ,\\
    & \eta_{S_1^-}\equiv \frac{1}{2}(\eta_{S_1}-\eta_{S_1^*})\, ,
\end{align}
such that
\begin{align}
    zHn_\gamma\frac{d\eta_{S_1^+}}{dz}=-\frac{1}{2}\left(D+\overline{D}+D_0+\overline{D}_0\right)\, \label{eq:sp1},\\
    zHn_\gamma\frac{d\eta_{S_1^-}}{dz}=-\frac{1}{2}\left(D-\overline{D}+D_0-\overline{D}_0\right)\, \label{eq:sm1}.
\end{align}
Next, we use chemical potential relations to rewrite the number densities of the light particles $H$, $\tilde R_2^\dagger$, $Q$, and $L$ in terms of the $B-L$ number density. To do this we begin by writing the number density $n_a$ of a particle $a$ in thermal equilibrium as~\cite{Harvey:1990qw}
\begin{equation}
\label{eq:num2chem}
   n_{a}= \frac{g_a T^3}{\pi^2}\times\left\{\begin{array}{ll}
         \displaystyle \zeta(3)+\frac{\mu_a}{T}\zeta(2) + \dots & \text{ (boson)}\\[3mm]
         \displaystyle \frac{3}{4}\zeta(3)+\frac{\mu_a}{T}\frac{\zeta(2)}{2} + \dots & \text{ (fermion)}
    \end{array}\right.\, ,
\end{equation}
where $\mu_a$ is the chemical potential, which we have assumed to be small, $\mu_a/T\ll 1$. Note that we have the relation $\mu_a=-\mu_{\bar a}$ for antiparticle $\bar a$. 
We now wish to relate the different chemical potentials to each other using the fact that some interactions are in chemical equilibrium in the early Universe. In the SM we have the three Yukawa interactions
\begin{equation}
\label{eq:chyuk}
    \mu_{q_i} = \mu_{\bar{u}^c_i}-\mu_H\, ,\qquad\mu_{q_i} = \mu_{\bar{d}^c_i}+\mu_H\, ,\qquad\mu_{L_i} = \mu_{\bar{e}^c_i}+\mu_H\, ,\qquad
\end{equation}
and electroweak sphaleron
\begin{equation}
\label{eq:chspah}
    \sum_i\mu_{q_i}=-\frac{1}{3}\sum_i\mu_{L_i}\, ,
\end{equation}
where $i$ denotes the generation. For the benchmark scenario chosen in Sec.~\ref{sec:results} using the leptoquark model presented in Sec.~\ref{sec:model} we can also assume that the reaction $\tilde R_2^\dagger \leftrightarrow L + \bar d$ is in equilibrium, which lets us write
\begin{equation}
\label{eq:chemr2}
    \mu_{\tilde R_2^\dagger} = \mu_{L_i}-\mu_{\bar{d}^c_i}\, .
\end{equation}
Further requiring a vanishing total $U(1)_Y$ hypercharge lets us write
\begin{equation}
    0=\sum_i Q_i^{U(1)_Y}(n_i-\bar n_i)\, ,
\end{equation}
where $Q_i^{U(1)_Y}$ is the $U(1)_Y$ charge of particle $i$, and the sum goes over all particle species that are in thermal equilibrium. 
We then relate the different chemical potentials to that of $\tilde R_2^\dagger$,
\begin{equation}
\label{eq:muR}
    \mu_a \equiv x_a\mu_{\tilde R_2^\dagger}\, .
\end{equation}
Next, we want to express $\mu_{\tilde R_2}$ in terms of the $B-L$ number density $n_{\Delta (B-L)}$ given by
\begin{equation}
    n_{\Delta (B-L)}\equiv n_{B-L}- n_{\bar B-\bar L}=\sum_i Q_i^{B-L}(n_i-\bar n_i)\, ,
\end{equation}
and we further relate $\mu_{\tilde R_2^\dagger}$ to the $B-L$ density
\begin{equation}
\label{eq:cBL}
    \frac{\eta_{\Delta (B-L)}}{\eta^\eq_{B-L}}\equiv C_{\tilde R_2^\dagger}\frac{\zeta(2)}{\zeta(3)}\frac{\mu_{\tilde R_2^\dagger}}{T}\, ,
\end{equation}
where explicitly we have
\begin{equation}
    C_{\tilde R_2^\dagger}=\frac{192-(12-68x_L)N}{96-9N}\, ,
\end{equation}
where $N$ is the number of fermion families. 
The Boltzmann equations in \cref{eq:sp1,eq:sm1} can now be written as
\begin{align}
   & zHn_\gamma\frac{d\eta_{S_1^+}}{dz}=-\gamma_D\left[\frac{\eta_{S_1^+}}{\eta_{S_1}^\eq}-1+\frac{\epsilon}{2}\frac{x_H+1+\frac{2}{3}(x_L+x_Q)}{C_{\tilde R_2^\dagger}}\frac{\eta_{\Delta (B-L)}}{\eta^\eq_{B-L}}\right]\, \label{eq:sp2},\\
   & zHn_\gamma\frac{d\eta_{S_1^-}}{dz}=-\gamma_D\left[\frac{\eta_{S_1^-}}{\eta_{S_1}^\eq}-\frac{\frac{r}{2}(x_H+1)-\frac{1}{3}(2-r)(x_L+x_Q)}{C_{\tilde R_2^\dagger}}\frac{\eta_{\Delta (B-L)}}{\eta^\eq_{B-L}}\right]\, \label{eq:sm2}.
\end{align}
The corresponding equation for $\eta_{\Delta (B-L)}$ can be expressed in terms of $\eta_{\tilde R_2^\dagger}$ using Eqs.~\eqref{eq:num2chem} and~\eqref{eq:cBL} such that
\begin{equation}
\label{eq:secondBE}
    zHn_\gamma\frac{\eta_{\Delta (B-L)}}{dz} = \frac{\eta^\eq_{B-L}}{2\eta_{\tilde R_2^\dagger}^\eq}C_{\tilde R_2^\dagger} zH n_{\gamma} \left(\frac{d\eta_{\tilde R_2^\dagger}}{dz}-\frac{d\eta_{\tilde R_2}}{dz}\right)\, .
\end{equation}
The equations for $\tilde R_2^\dagger$ and $\tilde R_2$ are in turn given by
\begin{align}
    & zHn_\gamma\frac{d\eta_{\tilde R_2^\dagger}}{dz}= D-S-2T_A-2T_B\\
    & zHn_\gamma\frac{d\eta_{\tilde R_2}}{dz}= \overline{D}-\overline{S}-2\overline{T_A}-2\overline{T_B}\, .
\end{align}
Here $S$ and $\overline S$ correspond to $s$-channel two-to-two scattering processes, and $T_{i}$ and $\overline{T_{i}}$ for $i\in\{A,B\}$ similarly correspond to $t$-channel diagrams (c.f.\ Fig.~\ref{fig:diagrams}), where the factor 2 accounts for the $u$-channel. The $s$-channel terms are given by
\begin{align}
    & S\equiv \gamma^\eq_\text{sub}(H\tilde R_2^\dagger\to\bar L\bar Q)\frac{\eta_H\eta_{\tilde R_2^\dagger}}{\eta_H^\eq\eta_{\tilde R_2^\dagger}^\eq}-\gamma^\eq_\text{sub}(\bar L\bar Q\to H\tilde R_2^\dagger)\frac{\eta_{\bar L}\eta_{\bar Q}}{\eta_L^\eq\eta_Q^\eq}\, ,\\
    & \overline S\equiv \gamma^\eq_\text{sub}(H^\dagger\tilde R_2\to L Q)\frac{\eta_{H^\dagger}\eta_{\tilde R_2}}{\eta_H^\eq\eta_{\tilde R_2^\dagger}^\eq}-\gamma^\eq_\text{sub}(L q\to H^\dagger\tilde R_2)\frac{\eta_{L}\eta_{Q}}{\eta_L^\eq\eta_Q^\eq}\, ,
\end{align}
where $\gamma^\eq_\text{sub}$ is the $s$-channel reaction rate with the on-shell contribution subtracted
\begin{equation}
    \gamma^\eq_\text{sub}(X a \to i j)\equiv \gamma^\eq(X a \to i j)-\gamma^\eq_\text{on-shell}(X a \to i j)\, .
\end{equation}
This subtraction is done in order to avoid double counting of the decay- and inverse decay processes. The relevant on-shell contributions can be written in terms of the decay reaction rate, e.g.\
\begin{equation}
\begin{aligned}
    \gamma^\eq_\text{on-shell}(H\tilde R_2^\dagger\to\bar L\bar Q) =& \gamma^\eq(H\tilde R_2^\dagger\to S_1)\BR{S_1\to \bar L\bar Q}\\
    =& \gamma^\eq(S_1^*\to H^\dagger\tilde R_2)\BR{S_1\to \bar L\bar Q}\, ,
\end{aligned} 
\end{equation} 
where in the last step we used $CPT$ invariance. We then find
\begin{align} 
    & \gamma^\eq_\text{on-shell}(H\tilde R_2^\dagger\to\bar L\bar Q) \approx \frac{\gamma_D}{2}\left(r-\frac{r^2}{2}-\epsilon\right)\, ,\\
    & \gamma^\eq_\text{on-shell}(\bar L\bar Q\to H\tilde R_2^\dagger) \approx \frac{\gamma_D}{2}\left(r-\frac{r^2}{2}+\epsilon\right)\, ,\\
    & \gamma^\eq_\text{on-shell}(H^\dagger\tilde R_2\to L Q) \approx \frac{\gamma_D}{2}\left(r-\frac{r^2}{2}+\epsilon\right)\, ,\\
    & \gamma^\eq_\text{on-shell}(L q\to H^\dagger\tilde R_2) \approx \frac{\gamma_D}{2}\left(r-\frac{r^2}{2}-\epsilon\right)\, .
\end{align} 
Here we have included $CP$-violating effects in the on-shell component that we subtract up to linear order in $\epsilon$, in order to remain consistent with the decay rate relations. However, we may safely neglect $CP$-violation in the full reaction rate without encountering inconsistencies, which we will do for simplicity by defining
\begin{equation}
    \gamma_S \equiv \gamma^\eq(H\tilde R_2^\dagger\to\bar L\bar Q) = \gamma^\eq(H^\dagger\tilde R_2\to L Q)\, .
\end{equation}
This leads to
\begin{align}
    & S = \left[\gamma_S-\frac{\gamma_D}{2}\left(r-\frac{r^2}{2}-\epsilon\right)\right]\frac{\eta_H\eta_{\tilde R_2^\dagger}}{\eta_H^\eq\eta_{\tilde R_2^\dagger}^\eq} - \left[\gamma_S-\frac{\gamma_D}{2}\left(r-\frac{r^2}{2}-\epsilon\right)\right]\frac{\eta_{\bar L}\eta_{\bar Q}}{\eta_L^\eq\eta_Q^\eq}\, ,\\
    & \overline S = \left[\gamma_S-\frac{\gamma_D}{2}\left(r-\frac{r^2}{2}-\epsilon\right)\right]\frac{\eta_{H^\dagger}\eta_{\tilde R_2}}{\eta_H^\eq\eta_{\tilde R_2^\dagger}^\eq} - \left[\gamma_S-\frac{\gamma_D}{2}\left(r-\frac{r^2}{2}-\epsilon\right)\right]\frac{\eta_{L}\eta_{Q}}{\eta_L^\eq\eta_Q^\eq}\, .
\end{align}
For the $t$-channel terms we have
\begin{align}
    &T_A = \gamma_{T_A}\left(\frac{\eta_{\tilde R_2^\dagger}\eta_Q}{\eta_{\tilde R_2^\dagger}^\eq\eta_Q^\eq}-\frac{\eta_{H^\dagger}\eta_{\bar L}}{\eta_H^\eq\eta_L^\eq}\right)\, ,\\
    &T_B = \gamma_{T_B}\left(\frac{\eta_{\tilde R_2^\dagger}\eta_L}{\eta_{\tilde R_2^\dagger}^\eq\eta_L^\eq}-\frac{\eta_{H^\dagger}\eta_{\bar Q}}{\eta_H^\eq\eta_Q^\eq}\right)\, ,\\
    &\overline{T_A} = \gamma_{T_A}\left(\frac{\eta_{\tilde R_2}\eta_{\bar Q}}{\eta_{\tilde R_2^\dagger}^\eq\eta_Q^\eq}-\frac{\eta_{H}\eta_{L}}{\eta_H^\eq\eta_L^\eq}\right)\, ,\\
    &\overline{T_B} = \gamma_{T_B}\left(\frac{\eta_{\tilde R_2}\eta_{\bar L}}{\eta_{\tilde R_2^\dagger}^\eq\eta_L^\eq}-\frac{\eta_{H}\eta_{Q}}{\eta_H^\eq\eta_Q^\eq}\right)\, ,
\end{align}
where we have again assumed $CP$-conservation for simplicity, such that
\begin{align}
   & \gamma_{T_A}\equiv \gamma^\eq(\tilde R_2^\dagger q \to H^\dagger \bar L) = \gamma^\eq(\tilde R_2\bar Q \to H L)\, ,\\
   & \gamma_{T_B}\equiv \gamma^\eq(\tilde R_2^\dagger L \to H^\dagger \bar Q) = \gamma^\eq(\tilde R_2\bar L \to H Q)\, .
\end{align}
We now find
\begin{equation}
\begin{aligned}
   zHn_\gamma\frac{d\eta_{\Delta (B-L)}}{dz} &=  \frac{\eta^\eq_{B-L}}{\eta_{\tilde R_2^\dagger}^\eq}\Bigg\{ C_{\tilde R_2^\dagger}\frac{\gamma_D}{2}\left[r\frac{\eta_{S_1^-}}{\eta_{S_1}^\eq}+\epsilon\left(\frac{\eta_{S_1^+}}{\eta_{S_1}^\eq}-1\right)\right]\\
    &-r\frac{\gamma_D}{2}\left[\frac{r}{2}(x_H+1)-\frac{1}{3}(2-r)(x_L+x_Q)\right]\frac{\eta_{\Delta (B-L)}}{\eta_{B-L}^\eq}\\
    & - \left(x_H+1+\frac{2}{3}(x_L+x_Q)\right)(\gamma_S+2\gamma_{T_A}+2\gamma_{T_B})\frac{\eta_{\Delta (B-L)}}{\eta_{B-L}^\eq}\Bigg\}\, .
\end{aligned}
\end{equation}
Using the chemical potential relations we see that we only need to find $x_L$, which is given by
\begin{equation}
    x_L=\frac{3N}{6N+2}\, .
\end{equation}
This lets us write our set of three Boltzmann equations as
\begin{align}
   zHn_\gamma\frac{d\eta_{S_1^+}}{dz} &=-\gamma_D\left[\frac{\eta_{S_1^+}}{\eta_{S_1}^\eq}-1+\epsilon\frac{1-\frac{4}{9}x_L}{C_{\tilde R_2^\dagger}}\frac{\eta_{\Delta (B-L)}}{\eta^\eq_{B-L}}\right]\, \label{eq:sp3},\\
   zHn_\gamma\frac{d\eta_{S_1^-}}{dz} &=-\gamma_D\left[\frac{\eta_{S_1^-}}{\eta_{S_1}^\eq}-\frac{r-\frac{4}{9}x_L(1+r)}{C_{\tilde R_2^\dagger}}\frac{\eta_{\Delta (B-L)}}{\eta^\eq_{B-L}}\right]\, \label{eq:sm3},
\end{align}
and
\begin{equation}
\begin{aligned}
   zHn_\gamma\frac{d\eta_{\Delta (B-L)}}{dz} &= \frac{\eta^\eq_{B-L}}{\eta_{\tilde R_2^\dagger}^\eq}\Bigg\{C_{\tilde R_2^\dagger}\frac{\gamma_D}{2}\left[r\frac{\eta_{S_1^-}}{\eta_{S_1}^\eq}+\epsilon\left(\frac{\eta_{S_1^+}}{\eta_{S_1}^\eq}-1\right)\right]\\
   &-\left[r\frac{\gamma_D}{2}\left(r-\frac{4}{9}x_L(1+r)\right)+\gamma_W\left(1-\frac{4}{9}x_L\right)\right]\frac{\eta_{\Delta (B-L)}}{\eta_{B-L}^\eq}\Bigg\}\, ,
   \label{eq:BE2f}
\end{aligned}
\end{equation}
where
\begin{equation}
    \gamma_W\equiv\gamma_S+2\gamma_{T_A}+2\gamma_{T_B}\, .
\end{equation}
Defining 
\begin{equation}
    c_+\equiv \frac{1-\frac{4}{9}x_L}{C_{\tilde R_2^\dagger}}\, ,\quad c_-\equiv \frac{r-\frac{4}{9}x_L(1+r)}{C_{\tilde R_2^\dagger}}\, ,\quad  c_{\Delta(B-L)}\equiv \frac{\eta^\eq_{B-L}}{\eta_{\tilde R_2^\dagger}^\eq}C_{\tilde R_2^\dagger}\, ,
\end{equation}
we finally have
    \begin{align}
    zHn_\gamma\frac{d\eta_{S_1^+}}{dz} &=-\gamma_D\left[\frac{\eta_{S_1^+}}{\eta_{S_1}^\eq}-1+\epsilon c_+\frac{\eta_{\Delta (B-L)}}{\eta^\eq_{B-L}}\right]\, ,\\
   zHn_\gamma\frac{d\eta_{S_1^-}}{dz} &=-\gamma_D\left[\frac{\eta_{S_1^-}}{\eta_{S_1}^\eq}-c_-\frac{\eta_{\Delta (B-L)}}{\eta^\eq_{B-L}}\right]\, ,\\
        \frac{zHn_\gamma}{c_{\Delta(B-L)}}\frac{d\eta_{\Delta (B-L)}}{dz} &= \frac{\gamma_D}{2}\left[r\frac{\eta_{S_1^-}}{\eta_{S_1}^\eq}+\epsilon\left(\frac{\eta_{S_1^+}}{\eta_{S_1}^\eq}-1\right)\right]-\left(c_-r\frac{\gamma_D}{2}+c_+\gamma_W\right)\frac{\eta_{\Delta (B-L)}}{\eta_{B-L}^\eq}\, .
    \end{align}

\bibliographystyle{JHEP}
\bibliography{References}
\end{document}